\documentclass[12pt,1p]{elsarticle}
\usepackage{graphicx}
\usepackage{amsmath,amssymb,amsthm}
\usepackage{hyperref}
\biboptions{longnamesfirst,semicolon}

\journal{Geometry and Physics}

\begin{document}

\begin{frontmatter}
\title{Using Cosmic Strings to Relate Local Geometry to Spatial Topology}
\author{Christopher Levi Duston}
\address{Physics Department, Merrimack College, N Andover, MA, 01845}
\ead{dustonc@merrimack.edu}


\newtheorem{theorem}{Theorem}
\newtheorem{definition}{Definition}
\newtheorem{lemma}{Lemma}
\newtheorem{prop}{Proposition}

\begin{abstract}
In this paper we will discuss how cosmic strings can be used to bridge the gap between the local geometry of our spacetime model and the global topology.  The primary tool is the theory of foliations and surfaces, and together with observational constraints we can isolate several possibilities for the topology of the spatial section of the observable universe. This implies that the discovery of cosmic strings would not just be significant for an understanding of structure formation in the early universe, but also for the global properties of the spacetime model. 
\end{abstract}
\begin{keyword}
spatial topology \sep cosmic strings \sep foliations \sep mathematical physics
\PACS 02.40.Vh \sep 04.20.Gz
\end{keyword}
\end{frontmatter}



\section{Introduction}
It is difficult to gather information on the global structure of the universe, primarily because a common feature of most physical theories is that they are local. In general relativity, this is explicit because the construction of a smooth manifold includes only a local definition of the metric. Combining relativity with cosmological observations provides a powerful tool to study the geometric properties of spacetime, but it does not give us much insight into the topological structure of the model. The primary signature of non-trivial topology is repeated patterns or ``ghost'' images in deep sky surveys \cite{Levin-etal-1998,Lachieze-Rey-Luminet-1995}, and the theoretical considerations for such structure are ad hoc.

Cosmic strings present a unique opportunity to study the topological nature of the universe because they provide a connection between the large scale structure and the local geometry. The underlying reason for this connection is that these strings are topological defects which have a well-defined local description. In this paper we will demonstrate that local observations of these defects can be used to classify the foliations allowed in the spacetime model, which can restrict the topological structure of the universe. 

\section{Cosmic Strings and Surfaces}\label{s:Cosmic_String_Review}
In this section we will review some basic facts about cosmic strings and isolate the context in which we are considering them. For more extensive reviews, see \cite{Vilenkin-Shellard-1994,Hindmarsh-Kibble-1995}.

A cosmic string is a byproduct of symmetry breaking in the early universe. Depending on the symmetry which is being broken, there can be ``defects'' in the field, and a cosmic string is such a 1-dimensional defect. More specifically, if the fields in question have a symmetry group $G$ that is broken due to a nonzero vacuum expectation value, then the symmetry group of the equivalent vacuua is $\mathcal{M}=G/H$, where $H$ is the unbroken subgroup. If this group is not simply-connected ($\pi_1(\mathcal{M})\neq 0$), then there are nontrivial loops of equivalent vacuua in the state space. If we assume $\mathcal{M}$ is path-connected, let us call such a loop $\gamma:[0,1]\to \mathcal{M}$ and $\gamma(0)=\gamma(1)$. Then for a nonzero vacuum state $\phi_0\neq 0$, we have $\gamma(x) \phi_0=\phi_0$ for all $x\in[0,1]$.

If there is a loop in physical space, $\omega:[0,1]\to M$ with $\omega(0)=\omega(1)$, then the vacuum of the field at each point in the loop is $\phi_0(\omega(y))$, $y\in [0,1]$. Traveling around this loop in space corresponds to traveling around some loop (possibly more than once) in $\mathcal{M}$ (otherwise the vacuum would not be unique); call this loop $\gamma$ from above. Then we will have 
\begin{equation}\gamma(x)\phi_0(\omega(y))=\phi_0(\omega(0)).\end{equation}
In other words, for every point on the curve $\omega$ there is an element in $\mathcal{M}$ that will return to $\phi_0$. Then, by assuming path-connectedness, we can shrink $\omega$ and $\gamma$ together. If the class of $[\gamma]\in\pi_1(\mathcal{M})$ is zero, then we can shrink them both to a point. If $[\gamma]\neq 0$, then the loop is not trivial and cannot be shrunk to a point, and the same applies to $\omega$. By playing this game with every closed loop in $M$, we can see that this develops into a 1-dimensional defect with points at the center of every loop we pick which localizes the string. If we were to consider the higher homotopy groups, we would find other topological defects such as monopoles or domain walls.

Physically, the string corresponds to a region of excess energy density where the original symmetry $G$ is unbroken. Because of the excess energy density, this string will cause gravitational distortions in the surrounding manifold. If $\mu$ is the energy density of the string and we take the limit of vanishing string thickness (usually called the \textit{straight string approximation} where $G\mu<<1$), we get a metric
\begin{equation} d s^2=- d t^2+ d z^2+ d r^2+(1-8G\mu)r^2 d \theta^2.\end{equation}
This is identical to a flat cylindrical metric under a coordinate transformation $\theta'=(1-4G\mu)\theta$, where $0\leq \theta'\leq 2\pi(1-4G\mu)$. The angular coordinate therefore has an ``angle deficit''
\begin{equation}\label{eq:AngleDeficit_Physics}
\Delta = 8\pi G\mu,
\end{equation}
and the constant $(t,z)$ surfaces have the geometry of cones. It is important to note that this is a purely local representation for the surface transverse to the string. Also note that the energy density and the string tension are related via \cite{Zwiebach-2004}
\begin{equation}\mu=\frac{T}{c^2},\end{equation}
so in units $c=1$, $\mu=T$ so the density and tension of the strings are interchangeable concepts.

Now consider a connected surface $\mathcal{S}$ which intersects a number of strings. These intersection points will generically cause the surface to be conical around the intersections. If this surface is compact, it can be given a flat metric as long as we allow for conical points \cite{Schumacher-Trapani-2005, Zorich-2006} (see also \ref{s:flatsurfaces}), so by identifying the conical points with the location of the strings, we can ensure our surfaces will be flat. Each conical point $p_i$ has an associated maximum value of the angular coordinate $\theta_i=2\pi(1-4G\mu_i)$. It is common to define $\theta_i=2 \pi(\beta_i+1)$, which gives an alternate definition for the angle deficit 
\begin{equation}\label{eq:AngleDeficit_Math}
\beta_i=-4G\mu_i
\end{equation}
for the tension $\mu_i$ of the string. The definition (\ref{eq:AngleDeficit_Physics}) is common in the physics literature, whereas (\ref{eq:AngleDeficit_Math}) is more common in the mathematics literature.

The scalar curvature of a flat compact surface with $m$ conical points is \cite{Fursaev-Solodukhin-1995}
\[R_{\mathcal{S}}=R_0-4\pi \sum_{i=1}^m \beta_i\delta(p_i),\]
where $\delta(p)$ is the delta function with support at $p$ and $R_0$ is the scalar curvature of the surface with all the points removed (that is, $R_0=0$). By integrating this expression and using the Gauss-Bonnet theorem, we see a global restriction on the sum of the angle deficits, which depends on the Euler characteristic of the surface $\chi(\mathcal{S})$:

\begin{equation}\label{eq:Flat_Condition}
2\pi\chi(\mathcal{S})+\sum_i^m(\theta_i-2\pi)=0.
\end{equation}

Locally, these two descriptions are identical, so there is no physical difference between the presence of cosmic strings and the presence of conical defects in our spacetime model. A cosmic string will create conical defects along its length, whereas conical defects of embedded surfaces will be mathematically identical to the presence of cosmic strings. Of course, there may be differences in the manifestation of these phenomena - for instance, it is thought that cosmic strings are either infinite in length or closed loops, whereas there is no \textit{a priori} reason to disregard either isolated or finite sets of conical singularities. If the transverse surface is additionally taken to be compact, we have a topological restriction coming from the presence of the strings and condition \ref{eq:Flat_Condition}. It turns out that the existance of any compact surfaces in the foliation will be a key restriction, and we will consider both the open and compact case in the rest of our analysis.

\section{Observational Constraints}\label{s:Observational_Constraints}
Because strings carry energy density, they can influence both the large-scale evolution of the universe and the local distribution of matter. Strings in the early universe will influence the growth conditions for galaxies, and will leave an observational signature in the Cosmic Microwave Background (CMB).

The most basic way to estimate the energy density of cosmic strings is to consider the string width to be equal to the Compton wavelength of the underlying field \cite{Vilenkin-Shellard-1994, Brandenberger-2014}. This gives the estimate
\begin{equation}G\mu\sim \left(\frac{\eta}{m_{pl}}\right)^2\end{equation}
for the symmetry breaking scale $\eta$ and Planck mass $m_{pl}$. Using the grand unification scale $\eta\sim 10^{16}$ GeV, we get the basic estimate $G\mu\sim 10^{-6}$.

A model-independent probe can be found if one considers how cosmic strings should affect observations of the CMB. The strings create ``wakes'' in spacetime causing Doppler shifts, leading to temperature fluctuations in the CMB. This was originally modeled by \cite{Bouchet-etal-1988, Hindmarsh-1994}, and a comparison to COBE data found $G\mu \leq 2.0 \times 10^{-6}$ \cite{Bennett-etal-1992}. The Planck group recently improved this limit to $\leq 3.2\times 10^{-7}$ \cite{Planck-Strings-2013}.

As gravitating objects, cosmic strings could also have played a significant role in structure formation in the early universe. An analysis of WMAP data puts these limits at $\sim 0.5 \times 10^{-6}$ \cite{Urrestilla-etal-2011}, and including data from the South Pole Telescope breaks some parameter degeneracy to yield a more stringent bound, $G\mu <1.7\times 10^{-7}$ \cite{Dvorkin-etal-2011}. 

If the surfaces transverse to the cosmic strings are compact, we can use these observational bounds with (\ref{eq:Flat_Condition}) to tightly constrain their topology. Along with (\ref{eq:AngleDeficit_Physics}) we can relate the sum of the string tensions to the Euler characteristic of the surface:
\begin{equation}\chi(S)=8\pi\sum G\mu_i.\end{equation}
To complete this estimate we will need to know how many strings we expect to see in the observable universe. This can be estimated by dynamical simulations \cite{Brandenberger-1994}, and the typical answer is the number of strings in one horizon volume is less than 10. Combining this with the strongest limits for the string tension, we estimate the observational bound on the Euler characteristic to be
\begin{equation}\label{eq:EulerLimit}
\chi(S)< 10^{-6}.
\end{equation}
The Euler characteristic of a surface is related to its genus by $\chi=2-2g$, and since it must be an integer, \ref{eq:EulerLimit} represents observational evidence that if cosmic strings exist, the surfaces which are transverse to them have genus $g=1$, and so have the topology of a torus $\mathbb{S}^1\times\mathbb{S}^1$. The assumption of connectedness is obviously a key component of this approach, since at this level nothing would prohibit the surface from being a disjoint sum such as
\begin{equation}\mathcal{S}=\mathbb{S}^1\times\mathbb{S}^1\sqcup \mathbb{S}^2\sqcup ...\sqcup \mathbb{S}^2.\end{equation}

\section{Consequences for the Spatial Topology of the Observable Universe}
In this section we will review some basic facts from foliation theory and construct a foliation for the spatial section of the observable universe. We will simply focus on what will be needed to understand our results - for a more extensive review, see \cite{Tamura-1992}. We will then discuss how these results can be used to learn something about the topology of our spacetime model.

\begin{definition}
A \textbf{codimension-$q$ foliation} $\mathcal{F}$ is a smooth $m$-manifold $M$ together with a set of immersed submanifolds $\{L_i\}$ (called \textbf{leaves}) such that $\bigcup_i L_i=M$ and the coordinate charts $\{(\mathcal{U}_\alpha,\phi_\alpha)\}$ on $M$ satisfy
\begin{equation}\phi_\alpha\left.\right|_{L_i}:\mathcal{U}_{\alpha}\to \mathbb{R}^{m-q} \end{equation}
\end{definition}
The picture for a foliation is a book, where each page is a leaf. If the dimension of the leaves is not constant, the foliation is said to be \textit{singular}. There are many important examples of foliations in the mathematical literature, but the one most familiar to physicists will be the foliation of the universe into 3-spheres, usually denoted $\mathbb{S}^3\times \mathbb{R}$.

A necessary condition for the existence of a foliation is that it possesses a \textbf{distribution} $E$, which is a choice of a $(n-q)$-dimensional subspace $E_p\subset T_pM$ of the tangent space at every point $p\in M$. The Frobenius theorem tells us that the condition for a distribution to represent a foliation is that it must be integrable. A distribution $E$ is integrable if it is the tangent space to the submanifold at every point,
\begin{equation}T_p\iota (T_pL_i)=E_p,\quad \forall p\in M.\end{equation}
 
Since we are making the straight string approximation, we have a preferred orthogonal subspace near the strings. By extending this subspace to the entire leaf, we have an explicit splitting of the tangent bundle,
\begin{equation}TM=T\mathcal{F}\oplus T\mathcal{S},\end{equation}
which means our foliation is transverse. In addition, since the spatial section $\Sigma$ is a 3-manifold, $\chi(\Sigma)=0$, and it supports a nowhere vanishing vector field. Using this vector field, we can give each $L_i$ an orientation, and our foliation is \textbf{transversely orientable}. 

Since we have a strict topological condition which can be applied in the case of compact leaves, we will need to know under what conditions we can choose the leaves to be compact. In fact, codimension-1 foliations are very rigid, and a single compact leaf can be a very restrictive global condition. For example, a basic result is \cite{Tamura-1992}:

\begin{theorem}
If a smooth 3-manifold $M$ with finite fundamental group allows a codimension-1 $C^r$ foliation $\mathcal{F}$ with $r\geq 2$, then $\mathcal{F}$ must have a compact leaf.
\end{theorem}
This motivates splitting our considerations into three categories - no compact leaves, at least one compact leaf, and foliations with entirely compact leaves.

\subsection{No Compact Leaves}
The presence of compact leaves are restrictive enough that if we suppose they don't exist, we can still learn something about the topology of the spatial section. Using the theorem above, if the foliation does not have a single compact leaf, $\pi_1(\Sigma)$ is infinite. Specifically, the spatial section is not simply-connected. There are additional observational hints of this behavior in the WMAP data, in the form of a preferred direction in space \cite{Tegmark-etal-2003, Stevens-etal-1993}.

\subsection{At Least One Compact Leaf}
If the foliation has at least one compact leaf, the resulting topology depends on if the leaf in question intersects a string or not.

\subsubsection{Does Not Intersect}
In this case we can use Reeb stability, and the result depends on the topology of the leaf, rather than the topology of the spatial section.

\begin{theorem}\cite{Tamura-1992}
Let $M$ be a smooth, compact, connected manifold with a codimension-1 $C^r$ foliation $\mathcal{F}$ with $r\geq 2$ which has a compact leaf with finite fundamental group. Then every leaf of $\mathcal{F}$ is compact with finite fundamental group.
\end{theorem}
If the leaf does not intersect a string but has finite fundamental group, then every other leaf must also have finite fundamental group. However, the observational constraints from \S\ref{s:Observational_Constraints} require the topology of closed surfaces intersecting the strings to have have genus $g=1$ (\textit{e.g.} $\pi_1(\mathcal{S})$ infinite). Therefore Reeb stability removes the possibility of a single compact leaf with finite $\pi_1(\mathcal{S})$. It has nothing to say about a single compact leaf with infinite fundamental group, but if cosmic strings are distributed evenly in space, this compact leaf would have to be small (cosmologically) to avoid intersections.

\subsubsection{Does Intersect}
If the compact leaf does intersect the strings, by our previous argument this is a $g=1$ surface. However, if it is the \textit{only} compact leaf (or the other compact leaves are outside it, \textit{e.g.} contained in $M\setminus N$ for $\partial N = \mathcal{S}$), then the foliation must be open inside. This situation describes a \textit{Reeb component}, in which a torus $\mathbb{S}^1\times \mathbb{S}^1$ is foliated internally by planes $\mathbb{R}^2$. This case is particularly interesting, since any closed 3-manifold can be generated by Dehn surgery along a torus in a 3-sphere \cite{PS,Gompf-Stipsicz-1999}. This can be done with a single torus or with multiple tori, and this construction would be rigid due to the presence of the strings (more on this in the next subsection). Such tori can also be combined with knot surgery to describe the matter content of the universe via exotic smoothness \cite{Asselmeyer-Maluga-Rose-2012}. So at the current observational status of cosmic strings we cannot say anything further here, but if cosmic strings were actually discovered, we might be able to use their pattern on the sky to isolate the precise topology of the entire spatial 3-manifold, or confirm other models requiring embedded tori.

\subsection{All Compact Surfaces}
If every leaf in the foliation is compact, then as we have already argued, these surfaces must be of genus 1, and here we point out that this condition is a) stable and b) global. If we adopt the cosmological principle, then we cannot appeal to coincidence - if cosmic strings exist in the numbers which simulations predict, then the sum of their angle deficits must be zero. This is a local condition, in the sense that we are talking about the geometry near the string. However, $\sum G\mu$ cannot be far from zero away from strings either, or that would be easily seen in the CMB. By constraining the string tension, one can constrain the existence and type of conical singularities, since they have the same observational signature. Another way to look at this is if the topology of the transverse surfaces was to be anything other than $g=1$ (say $g=0$, topologically $\mathbb{S}^2$), say with conical singularities away from the strings that had negative angle deficits, we would observe that as a massive source for cosmic strings. The lack of this signature means that if cosmic strings exist, the surfaces transverse to these strings are tori.

This argument only applies directly to the \textit{observable} universe, but appealing again to the cosmological principle, we should be able to extend the foliation to the entire spatial section. We will also note that the existence such a foliation is not a surprise, given the fundamental work done by Thurston on this topic \cite{Thurston-1974}. He found that every closed connected manifold with zero Euler characteristic has a smooth codimension-1 foliation. This guarantees that we could find a foliation of the spatial section, but our goal here is to describe the specific foliation provided by the cosmic strings, and illustrate how it can be used to determine the topology of space.

\section{Summary}
We have proved the following:
\begin{theorem}
If cosmic strings exist, the straight string approximation from \S\ref{s:Cosmic_String_Review} is valid, and the observational constraints of \S\ref{s:Observational_Constraints} apply, a foliation $\mathcal{F}$ of the spatial section $\Sigma$ exists, and there are three possibilities for its topology:
\begin{itemize}
\item The foliation contains no compact surfaces, and $\Sigma$ is not simply-connected.
\item The foliation contains Reeb components corresponding to toroidal elements of a surgery.
\item The foliation contains a single, cosmologically small compact leaf and the topology of $\Sigma$ is otherwise unknown.
\end{itemize}
\end{theorem}
From an observational standpoint, this situation is very optimistic. Actual confirmation of cosmic strings would not only be important for high energy physics and cosmology, but it would also help to restrict the topology of the universe. Theoretically, the main interest here is the very existence of such bounds on topology of the spatial section. Since most of our physical theories are built on the local behavior of quantum fields, it is generically very difficult to access the global structure of spacetime. Cosmic strings provide a unique opportunity to study this global structure by connecting high energy theory and phenomenological cosmology with surface and foliation theory. It is our hope that this work will be just a first step, and it will reveal other opportunities to study the global structure of spacetime via local geometry, either from cosmic strings or other unique, unexpected phenomena.

\section*{Acknowledgments}
The author would like to thank Mark Brittenham for providing access to an excellent set of notes on foliations which initiated the early lines of inquiry in this paper.

\appendix
\section{Flat Surfaces}\label{s:flatsurfaces}
In this appendix we will briefly discuss how one can assign a flat metric to any surface, provided one allows for the existance of conical singularities. For more details, we refer to the reader to \cite{Strebel-1984,Masur-Tabachnikov-2002, Dankwart-2010}.

We will assume our surface $\mathcal{S}$ is Riemann with an atlas $\{(\mathcal{U}_i,z_i)\}$ where $z_i:\mathcal{U}\to \mathbb{C}$ are homeomorphisms and the overlaps, $z_i\circ z_j^{-1}:\mathcal{U}_i\bigcap\mathcal{U}_j\to \mathbb{C}$, are complex analytic. Any analytic function $\phi$ in an open set $V\subset \mathbb{C}$ defines a line element $dz$ by requiring that $\phi dz^2$ is real and positive. Specifically, $\text{arg}(dz)=-1/2~\text{arg}(\phi (z))+n\pi/2$ and $|dz|=\sqrt{|\phi (z)|}$. On the overlap $\mathcal{U}_i\bigcap\mathcal{U}_j$, $\phi$ has the transformation property
\[\phi_j(z_j)=\phi_i(z_i)\left(\frac{dz_i}{dz_j}\right)^2.\]
$\phi$ is called a \textit{quadratic differential}. We can use this analytic function to define a \textit{natural coordinate} on $\mathcal{U}_i\subset \mathcal{S}$,
\[w(z_i)=\int_{z_i(p)}^{z_i}\sqrt{\phi(\zeta)}d\zeta,\]
in a region of a regular point $p$ (\text{i.e.} not a zero or a pole). The transformation law for the quadratic differential in these coordinates is
\[dw^2=\phi(\zeta)d\zeta^2,\]
so in the natural coordinates we have $\phi(w)=1$.

We now assume that $\phi$ has only simple poles (the degree of the poles are greater or equal to -1). The norm of the quadratic differential
\[|\phi|=\int_\mathcal{S} |\phi(z)||dz^2|\]
is therefore finite and determines a metric with length element $\sqrt{|\phi(z)|}|dz|$. In the natural coordinates $w$ above, this metric will be flat.

\begin{figure}[h]\begin{center}
\includegraphics[scale=0.3]{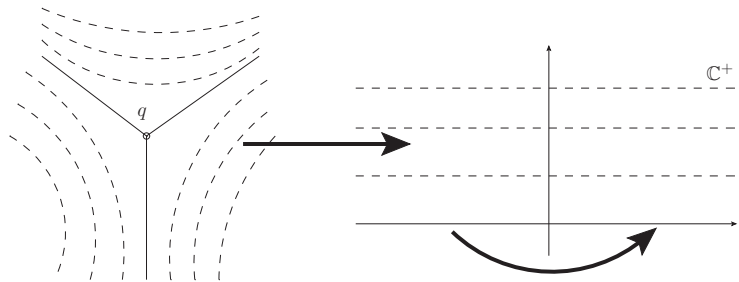}
\caption{The region around a pole of order 1, which maps to 3 copies of a half-plane with the identification $-1\sim 1$ in $\mathbb{C}$ under the natural coordinates.}\label{fig:kplus2regions}
\end{center}\end{figure}

In a region around a zero or simple pole $q$ the quadratic differential can be written
\[\phi(z)dz ^2=\left(\frac{n+2}{2}\right)^2z^ndz^2\]
where $z(q)=0$ (Theorem 6.2 in \cite{Strebel-1984}). In natural coordinates we can perform the integral above to find
\[w(z)=z^{\left(\frac{n+2}{2}\right)},\]
so $w$ maps the open region around $q$ to $n+2$ half-planes as shown in figure \ref{fig:kplus2regions}. The metric in this region can be written
\[ds^2=dr^2+(crd\theta)^2\]
with $c=(n+2)/2$, and so the order of $\phi$ around a cosmic string with string tension $\mu$ is
\[n=-16G\mu.\]
Therefore, to avoid poles  of order $n\leq -2$, we require
\[G\mu\leq \frac{1}{16},\]
which is easily satisfied by current observations (see \S\ref{s:Observational_Constraints}).

\section*{References}

\bibliography{CosmicStringsBib}

\end{document}